\documentstyle[twoside,fleqn,espcrc2,epsfig]{article}

% ...

\newcommand{\AmS}{{\protect\the\textfont2
 A\kern-.1667em\lower.5ex\hbox{M}\kern-.125emS}}

\title{Bragg- and Moving-glasses: a theory of disordered vortex lattices}

\author{T. Giamarchi
\address{Laboratoire de Physique des Solides,
Universit{\'e} Paris-Sud, B\^at 510, 91405 Orsay, France}
\thanks{Laboratoire associated to CNRS; email:
giam@lps.u-psud.fr}
and
P. Le Doussal\address{CNRS-Laboratoire de Physique Th\'eorique de l'Ecole
Normale Sup\'erieure, 24 rue Lhomond, F-75231 Paris, France}
\thanks{email: ledou@physique.ens.fr}}

\begin{document}

\begin{abstract}
We study periodic lattices,
such as vortex lattices in type II superconductors
in a random pinning potential.

For the static case we review the prediction
\cite{giamarchi_vortex_longshort,giamarchi_diagphas_prb}
that the phase diagram of such systems consists of a
topologically ordered Bragg glass phase, with quasi long range
translational order, at low fields. This Bragg glass phase undergoes a
transition at higher fields into another glassy phase, with
dislocations, or a liquid. This proposition is compatible with a large
number of experimental results on BSCCO or Thalium compounds. Further
experimental consequences of our results and relevance to other
systems will be discussed.

When such vortex systems are driven by an external force, we
show that, due to periodicity
in the direction transverse to motion, the effects of static disorder
persist even at large velocity\cite{giamarchi_moving_prl}.
In $d=3$, at weak disorder, or large velocity
the lattice forms a topologically ordered {\it glass}
state, the ``moving Bragg glass'',
an anisotropic version of the static Bragg glass.
The lattice flows through well-defined, elastically coupled, static
channels. We determine the roughness of the manifold
of channels and the positional correlation functions.
The channel structure also provides a natural starting
point to study the influence of topological defects such as
dislocations. In $d=2$ or at
strong disorder the channels can decouple along the direction of
motion leading to a ``smectic'' like flow.
We also show that such a structure exhibits
an effective transverse critical pinning force due to barriers to
transverse motion, and discuss the experimental consequences of this
effect.
\end{abstract}


\maketitle

\section{Introduction}

To understand the effects of static substrate disorder
on periodic media such as vortex lattices \cite{blatter_vortex_review}
is of paramount importance both for the
technological applications of high-$T_c$ materials and from a purely theoretical
point of vue. Such a study also applies to many other physical
problems such as charge density waves (CDW),
Wigner crystals, colloids,
magnetic bubbles.

For the vortex
lattices it is generally agreed
that disorder leads to a glass state with diverging
barriers and pinning
\cite{fisher_vortexglass_global,feigelman_collective}.
Although the precise nature of the
glass state has been the subject of much debate, most theories
agreed on the absence of translational order.
General arguments also tended to prove that
disorder would always favor the presence of
dislocations \cite{fisher_vortexglass_global}.
Many points were not naturally fitting in the framework
of these theories such as the existence of a first order transition
\cite{charalambous_melting_rc,safar_tricritical_prl},
decoration experiments of showing remarkably large regions
free of dislocations \cite{grier_decoration_manips}.
On the side of theory, scaling arguments
\cite{nattermann_pinning_etal} suggested, within a purely
elastic decription, a slower, logarithmic, growth of deformations.

In a recent work we have obtained the first quantitative theory of the
elastic vortex lattice
\cite{giamarchi_vortex_longshort} in presence
of point disorder \cite{korshunov_variational_short}.
Contrarily to
previous approaches, it provides a description valid at {\bf all scales}
and demonstrates
that while disorder produces algebraic growth of displacements at short
length scales,
periodicity takes over at large scales and results in a
decay of translational order {\it at most algebraic}.
Moreover we showed \cite{giamarchi_vortex_longshort} using energy
arguments that dislocations are {\it not favorable}
for weak disorder in $d=3$. This implies the existence
of a {\bf thermodynamic} glass phase, as far as energy and very low
current transport properties are
concerned, retaining a {\bf nearly perfect} (i.e. algebraic)
translational order and thus Bragg peaks: the ``Bragg glass''.
Because it retains a ``lattice'' structure
and Bragg peaks, this glass phase is radically different from the vortex
glass picture.
We proposed \cite{giamarchi_vortex_longshort} that the phase seen
experimentally at low fields was the Bragg glass, solving the
apparent impossibility of a pinned solid.
This accounted naturally for the first-order transition
and the decoration experiments.
Our prediction \cite{giamarchi_vortex_longshort}
received subsequent further support both from numerical
\cite{gingras_ryu_etal} or analytical
\cite{kierfeld_bglass_layered,carpentier_bglass_layered} calculations.

\begin{figure}
\label{figure1}
\centerline{\epsfig{file=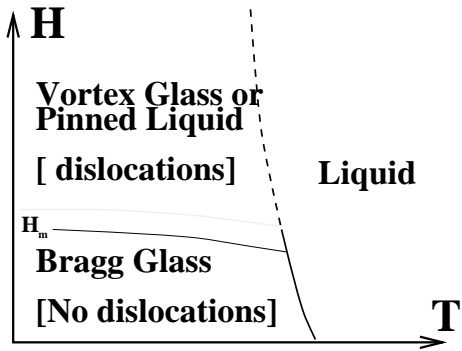,angle=-90,width=6cm}}
\caption{The stability region of the Bragg glass phase in the magnetic
field $H$, temperature $T$ plane is shown schematically. The thick line
is expected to be first order, whereas the dotted line should be either
second order or a crossover. Upon increasing
disorder the field induced melting occurs for lower fields as
indicated by the thin solid line. }
\end{figure}

Upon raising the field, or equivalently increasing the disorder, the
Bragg glass should disappear by spontaneous generation of dislocations,
when the translational length $R_a$ (the length for which relative
displacements become of order of the lattice spacing $a$) is of the
order of $a$. The critical point occuring on the melting line
\cite{zeldov_diagphas_bisco,safar_tricritical_prl}
was the end point of the transition line between the Bragg glass
at low fields
and a topologically disordered glassy phase (or a strongly
pinned liquid) at higher field. The topology of the phase diagram
proposed in \cite{giamarchi_vortex_longshort}
is as depicted in Fig.~\ref{figure1}.

Several recent experiments can be interpreted to confirm the picture
proposed in \cite{giamarchi_vortex_longshort}.
Neutron experiments can be naturally interpreted in term of
the Bragg glass \cite{yaron_neutrons_vortexetal}.
In BSCCO neutron peaks are observed at low fields and disappear upon
raising the field \cite{cubbit_neutrons_bscco}. The phase diagram of BSCCO
\cite{zeldov_diagphas_bisco} is also compatible with of our theory, the
second magnetization peak line corresponding to the predicted
field driven transition.
A numerical estimate \cite{giamarchi_diagphas_prb} of the melting field
$H_M$ for BiSCCO gives
$H_M\sim 400 G$ in good agreement with the observed experimental values
\cite{khaykovich_diagphas_bisco}. This line is found to be relatively
temperature independent
at lower temperatures and to be shifted downwards upon increase of point disorder
\cite{khaykovich_diagphas_bisco,chikumoto_secondpeak_bisco}.
Similar types of phase diagrams are also observed in a variety of
materials, including YBCO, organic superconductors and heavy fermion
compounds. More experimental consequences and references can be
found in \cite{giamarchi_diagphas_prb}.

Let us now turn to the dynamics of the driven system.
It is crucial to determine how much of the glassy properties
remain, and how translational
and topological order behave \cite{higgins_dynamics_phasediag}.
Indeed several experiments suggest
that a fast moving lattice is more ordered than a static one
\cite{thorel_neutron_vortex,yaron_neutrons_vortexetal}.
Arguments \cite{koshelev_dynamics} based on large velocity
expansion \cite{schmidt_hauger_etal} concluded
that at low $T$ and above a certain velocity
the moving lattice is a crystal at an effective temperature
$T^{\prime }=T+T_{sh}$, with bounded displacements and no glassy
properties.

We showed \cite{giamarchi_moving_prl} that in fact
the perturbation theory \cite{koshelev_dynamics}
breaks down, even at large $v$.  Some modes
of the disorder are not affected by the
motion and the periodic structure driven along $x$
experiences a transverse {\bf static} pinning force $F_{\rm stat}$,
perpendicular to the direction of motion.
\begin{equation}
F_\alpha ^{{\rm stat}}(r,u) = V(r)\rho_0\sum_{K.v=0} iK_\alpha \exp
(iK\cdot(r-u))
\end{equation}
where the random potential has correlations $\langle
V(r)V(r^{\prime})\rangle = \Delta(r-r^{\prime})$
of range $r_f$.
This force originates only from the
periodicity along $y$ and the uniform density modes along $x$
\cite{giamarchi_moving_prl,giamarchi_reply_movglass}.
It leads to a {\bf moving glass} (MG).

\begin{figure}
\centerline{\epsfig{file=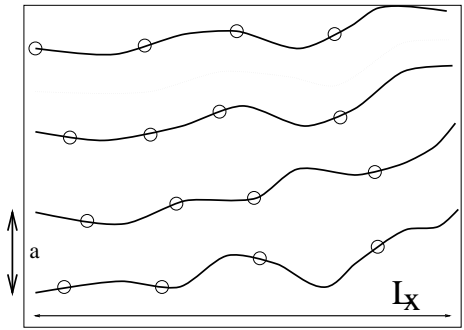,angle=-90,width=6cm}}
\caption{Motion in the moving glass occurs through static channels
wandering at distance $a$ over lengths $L_x \sim L_y^2$. If
dislocations are present ($d=2$ or strong disorder in $d=3$) they
should lead to a decoupling of channels, as indicated by the dotted
line.}
\label{channels}
\end{figure}

\begin{figure}
\centerline{\epsfig{file=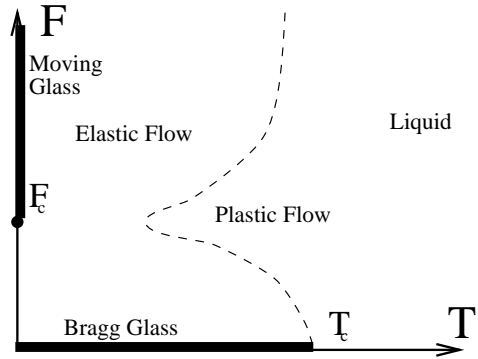,angle=-90,width=6cm}}
\caption{Phase diagram in force $F$, temperature $T$, for weak disorder
and in $d=3$. At zero
external force the system is in the free Bragg glass state. At large
velocities, in the moving Bragg glass one. This suggests that in this
case the depinning transition could be purely elastic.}
\label{ftdiag}
\end{figure}

We analyse here this result more quantitatively, using
a functional RG to study the equation of motion.
We only quote
here the RG equation for the disorder term, for the
periodic structure at $T=0$ (for the complete calculation see
\cite{giamarchi_mglass_long})
\begin{equation} \label{shortequ}
\frac{ d \Delta(u)}{dl}
= \Delta(u) +  \Delta''(u) ( \Delta(0)  - \Delta(u) )
\end{equation}
where a factor $\frac{1}{4 \pi v c} \epsilon$, with $\epsilon=3-d$
has been absorbed in $\Delta(u)$ (chosen to be of period $1$).
For $d>3$ disorder renormalizes to zero and the moving system is a
crystal. For $d < 3$ $\Delta$ flows to a new fixed point
$\Delta^{*}(u) = \Delta(0)(l) + u^2/2 - u/2$,
showing that the static disorder is still relevant in the moving
structure
(with the same conclusion in $d=3$,\cite{giamarchi_mglass_long}).
The value of $\Delta(0)(l)$ grows unboundedly as
$\Delta(0)(l) = \Delta(0) e^{\epsilon l}$ which indicates the existence
of a random force along the $y$ direction, generated under
renormalization. A similar force is generated along $x$
\cite{giamarchi_mglass_long}.
This does not spoil the above fixed point, since one can always
separates $\Delta(0)$ and  $\Delta(u) - \Delta(0)$.
This has several consequences. In particular, for $d \leq 3$ the system
remains a glass and motion occurs through elastic channels, as shown on
figure~\ref{channels}.
They are the easiest paths where particle follow
each other in their motion. They form a manifold
of elastically coupled, almost parallel lines or sheets (for
vortex lines in $d=3$) directed along $x$
and characterized by some transverse wandering $u_y$.
In the laboratory frame they are determined by the static disorder
and do not fluctuate with time.
Such channels were subsequently observed in numerical simulations
\cite{moon_moving_numerics} and in recent decoration in motion
experiments \cite{marchevsky_decoration_channels}.
In $d=3$ diplacements only grow logarithmicaly,
so the MG conserves quasi-long range translational order.
Thus similarly to the statics, the MG in $d=3$ at
weak disorder or large velocity is expected to retain perfect
topological order. This leads to the $F-T$ phase diagram of
figure~\ref{ftdiag}.
In $d=2$ however displacements grow algebraically and dislocations are
likely to appear. The existence of channels \cite{giamarchi_moving_prl}
then {\it naturally} suggests
a scenario by which dislocations affect the MG: when
the periodicity along $x$ is retained, e.g., presumably in
$d=3$ at weak disorder, the channels are coupled along $x$.
Upon increasing disorder or decreasing velocity in $d=3$,
or in $d=2$, decoupling between channels can occur,
reminiscent of static decoupling in a layered geometry
\cite{carpentier_bglass_layered}.
Dislocations are then inserted between the layers,
naturally leading to a flowing smectic glassy state,
recently observed in $d=2$ numerical simulations
\cite{moon_moving_numerics}.
Indeed, the transverse smectic order is likely to be more
stable than topological order along x, because of particle
conservation \cite{giamarchi_mglass_long}.

As an important consequence of the existence of the MG,
barriers for transverse motion exist
once the pattern of channels is established. Thus the
response to an additional small transverse force $F_y$
is very non linear with activated behaviour. At $T=0$
and neglecting the dynamic part of the disorder a true
transverse critical current $J_y^c$ exists.
This pinning force can directly be obtained from the RG equation
(\ref{shortequ}). Since its fixed point has
a {\it non analycity} at $u=0$, (leading to
$\Delta'(0^+) = 1/2$) there is a critical force,
determined at the Larkin length $L_y$:
\begin{equation} \label{critforce}
F_c = \int dq G(q) \Delta'(0^+) =
\frac{\epsilon}{4 \pi v c_y}
\sim L_y^{-2}
\end{equation}
(\ref{critforce}) coincides with the more qualitative derivation
consisting in  balancing the pinning energy
with the transverse Lorentz force acting on a Larkin domain
\cite{giamarchi_moving_prl}. Unlike in the $v=0$ case, the critical
force does not kill the random force in the FRG equation.
The MG is dominated by the
competition between the random force and the critical force.

These predictions
can be tested in experiments on the vortex lattice, or other systems
such as colloids, magnetic bubbles or CDW, or
or numerical simulations. Additional physical consequences and
references can be found in
\cite{giamarchi_moving_prl,giamarchi_mglass_long}.


\begin{thebibliography}{10}

\bibitem{giamarchi_vortex_longshort}
T. Giamarchi and P. {Le Doussal}, Phys. Rev. Lett. {\bf 72},  1530  (1994);
Phys. Rev. B {\bf 52},  1242  (1995).

\bibitem{giamarchi_diagphas_prb}
T. Giamarchi and P. {Le Doussal}, Phys. Rev. B {\bf 55},  6577  (1997).

\bibitem{giamarchi_moving_prl}
T. Giamarchi and P. {Le Doussal}, Phys. Rev. Lett. {\bf 76},  3408  (1996).

\bibitem{blatter_vortex_review}
G. Blatter {\it et~al.}, Rev. Mod. Phys. {\bf 66},  1125  (1994).

\bibitem{fisher_vortexglass_global}
M.~P.~A. Fisher, Phys. Rev. Lett. {\bf 62},  1415  (1989);
D.~S. Fisher, M.~P.~A. Fisher, and D.~A. Huse, Phys. Rev. B {\bf 43},  130
(1990).

\bibitem{feigelman_collective}
M. Feigelman, V.~B. Geshkenbein, A.~I. Larkin, and V. Vinokur, Phys. Rev. Lett.
  {\bf 63},  2303  (1989).

\bibitem{charalambous_melting_rc}
M. Charalambous, J. Chaussy, and P. Lejay, Phys. Rev. B {\bf 45},  5091
  (1992).

\bibitem{safar_tricritical_prl}
H. Safar {\it et~al.}, Phys. Rev. Lett. {\bf 70},  3800  (1993).

\bibitem{grier_decoration_manips}
D.~G. Grier {\it et~al.}, Phys. Rev. Lett. {\bf 66},  2270  (1991).

\bibitem{nattermann_pinning_etal}
T. Nattermann, Phys. Rev. Lett. {\bf 64},  2454  (1990);
J. Villain and J.~F. Fernandez, Z Phys. B {\bf 54},  139  (1984).

\bibitem{korshunov_variational_short}
S.~E. Korshunov, Phys. Rev. B {\bf 48},  3969  (1993).

\bibitem{gingras_ryu_etal}
M.~J.~P. Gingras and D.~A. Huse, Phys. Rev. B {\bf 53},  15193 (1996);
S. Ryu, A. Kapitulnik, and S. Doniach, Phys. Rev. Lett. {\bf 77},  2300  (1996) and
references therein.

\bibitem{kierfeld_bglass_layered}
J. Kierfeld, T. Nattermann, and T. Hwa, Phys. Rev. B {\bf 55}, 626
(1997).

\bibitem{carpentier_bglass_layered}
D. Carpentier, P. {Le Doussal}, and T. Giamarchi, Europhys. Lett. {\bf 35},
  379  (1996).

\bibitem{zeldov_diagphas_bisco}
E. Zeldov and al., Nature {\bf 375},  373  (1995).

\bibitem{yaron_neutrons_vortexetal}
U. Yaron and al., Phys. Rev. Lett. {\bf 73},  2748  (1994);
T. Giamarchi and P. {Le Doussal}, Phys. Rev. Lett. {\bf 75},  3372  (1995).

\bibitem{cubbit_neutrons_bscco}
R. Cubbit and al., Nature {\bf 365},  407  (1993).

\bibitem{khaykovich_diagphas_bisco}
B. Khaykovich and {et al.}, Phys. Rev. Lett. {\bf 76},  2555  (1996).

\bibitem{chikumoto_secondpeak_bisco}
N. Chikumoto and al., Physica C {\bf 185-189} 2201 (1991);
Phys. Rev. Lett. {\bf 69}, 1260  (1992).

\bibitem{higgins_dynamics_phasediag}
M. J. Higgins and S. Bhattacharya, Physica C {\bf 257}, 232 (1996).

\bibitem{thorel_neutron_vortex}
R. Thorel and al., J. Phys. (Paris) {\bf 34},  447  (1973).

\bibitem{koshelev_dynamics}
A.~E. Koshelev and V.~M. Vinokur, Phys. Rev. Lett. {\bf 73},  3580  (1994).

\bibitem{schmidt_hauger_etal}
A. Schmidt and W. Hauger, J. Low Temp. Phys {\bf 11},  667  (1973);
A.~I. Larkin and Y.~N. Ovchinnikov, Sov. Phys. JETP {\bf 38},  854  (1974).

\bibitem{giamarchi_reply_movglass}
T. Giamarchi and P. {Le Doussal}, Phys. Rev. Lett. {\bf 78},  752  (1997).

\bibitem{giamarchi_mglass_long}
T. Giamarchi and P. {Le Doussal}, 1997, to be published.

\bibitem{moon_moving_numerics}
K. Moon and al., Phys. Rev. Lett. {\bf 77},  2378  (1997),
S. Ryu et al. to be published.

\bibitem{marchevsky_decoration_channels}
M. Marchevsky and al., Phys. Rev. Lett. {\bf 78},  531  (1997).

\end{thebibliography}

\end{document}